%% file: gmin2-mw.tex
\pdfoutput=1
\documentclass[12pt]{article}
\usepackage{graphicx}
\usepackage{subcaption}
\usepackage{amssymb}
\usepackage{amsmath}
\usepackage{amsfonts}
\usepackage{multirow}

\usepackage{xcolor}
\usepackage{url}
\usepackage{ulem}
\usepackage{float}
\usepackage{float}

\usepackage{pifont}% http://ctan.org/pkg/pifont

\usepackage{soul}
\usepackage[english]{babel}
\usepackage[T1]{fontenc}
\usepackage{lmodern}
\usepackage[utf8]{inputenc}  % f??r Unix-Systeme
\usepackage{bbm}

\usepackage{hyperref}
\hypersetup{
  colorlinks   = true, %Colours links instead of ugly boxes
  urlcolor     = blue, %Colour for external hyperlinks
  linkcolor    = black, %Colour of internal links
  citecolor   = black %Colour of citations
}

\usepackage[numbers,sort&compress]{natbib}
\graphicspath{{plots/}}

\input{paperdef.tex}

\newcommand{\fnalbnlsig}{\htrs{0.8}}   % Difference FNAL and BNL [sigma]
\newcommand{\newcen}{\htrs{206.1}}     % new central value [10^-10]
\newcommand{\newunc}{\htrs{4.1}}       % new uncertainty [10^-10]
\newcommand{\newdiff}{\htrs{25.1}}    % new difference to SM [10^-10]
\newcommand{\newdiffunc}{\htrs{5.9}}   % new uncertainty of diff to SM [10^-10]
\newcommand{\newdiffsig}{\htrs{4.2}}   % new difference SM - exp [sigma]
\newcommand{\htrs}[1]{{\color{black} #1}}

\evensidemargin \oddsidemargin
\allowdisplaybreaks
\begin{document}
\thispagestyle{empty}

\def\thefootnote{\fnsymbol{footnote}}

\begin{flushright}
\mbox{}\
CERN--TH--2022--034,
DESY--22--041,
IFT--UAM/CSIC--22-029, %\\
%IPMU21-00II%\\
arXiv:2203.15710 [hep-ph]
\end{flushright}

\vspace{0.5cm}

\begin{center}

{\large\sc
{\bf Interdependence of the new ``MUON G-2'' Result\\[.3em]
and the \boldmath{$W$}-Boson Mass
}}

\vspace{0.5cm}

{\sc
Emanuele Bagnaschi$^{1}$%
%\footnote{email: Emanuele.Bagnaschi@cern.ch}%
, Manimala Chakraborti$^{2}$%
%\footnote{email: mani.chakraborti@gmail.com}%
, Sven Heinemeyer$^{3}$%
%\footnote{email: Sven.Heinemeyer@cern.ch}%
,\\[0.2em] Ipsita Saha$^{4}$%
%\footnote{email: ipsita.saha@ipmu.jp}%
~and Georg Weiglein$^{5,6}$%
\footnote{emails: Emanuele.Bagnaschi@cern.ch,
mani.chakraborti@gmail.com,
Sven.Heinemeyer@cern.ch,\\
\mbox{}\hspace{19mm}ipsita.saha@ipmu.jp,
Georg.Weiglein@desy.de}%
}

\vspace*{.7cm}

{\sl
$^1$CERN, Theoretical Physics Department, CH-1211 Geneva 23, Switzerland

\vspace{0.1cm}

$^2$Astrocent, Nicolaus Copernicus Astronomical Center of the Polish
Academy of Sciences, ul. Rektorska 4, 00-614 Warsaw, Poland

\vspace{0.1cm}

$^3$Instituto de F\'isica Te\'orica (UAM/CSIC),
Universidad Aut\'onoma de Madrid, \\
Cantoblanco, 28049, Madrid, Spain

\vspace{0.1cm}

$^4$Kavli IPMU (WPI), UTIAS, University of Tokyo, Kashiwa, Chiba 277-8583, Japan

\vspace{0.1cm}

$^5$Deutsches Elektronen-Synchrotron DESY,
Notkestr.~85, 22607 Hamburg, Germany

\vspace{0.1cm}

$^6$Universit\"at  Hamburg, Luruper Chaussee 149, 22761 Hamburg, Germany
}

\end{center}

\vspace*{0.1cm}

\begin{abstract}
\noindent
The electroweak (EW) sector of the Minimal Supersymmetric extension of the
Standard Model (MSSM), assuming the lightest neutralino as Dark Matter (DM)
candidate, can account for a variety of experimental results. This
includes the DM direct detection limits, the searches for EW
superpartners at the
LHC, and in particular the discrepancy between the
experimental result for the anomalous magnetic moment of the muon, \gmin2, and
its Standard Model (SM) prediction.
The new ``MUON G-2'' result, combined
with the older BNL result on \gmin2, yields
a deviation from the SM prediction of
$\De\amu = (\newdiff \pm \newdiffunc) \times 10^{-10}$, corresponding to
$\newdiffsig\,\sig$.
Using this updated bound,
together with the other constraints, we calculate the MSSM
prediction for the mass of the $W$~boson, $\MW$.
We assume contributions only from the EW
sector, i.e.\ the colored sector of the MSSM, in agreement with the
search limits, is taken to be heavy.
We investigate five scenarios, distinguished by the mechanisms which yield
a relic DM density in agreement with the latest Planck bounds.
We find that with the new \gmin2\ result taken into account and
depending on the scenario, values up to
$\MWMSSM \lsim 80.376 \gev$ are reached.
The largest values are obtained for wino DM and in the case of slepton
co-annihilation, where points well within
the $1\,\sigma$ range of the experimental world average of
$\MWexp = 80.379 \pm 0.012 \gev$ are reached,
whereas the SM predicts a too small value of $\MWSM = 80.353 \gev$.
We analyze the dependence of $\MWMSSM$ on the relevant masses of the EW
superpartners
and demonstrate that future $\MW$ measurements, e.g.\ at the ILC, could
distinguish between various MSSM realizations. Sizable contributions to
$\MWMSSM$ are associated with a relatively light $\neu1$, accompanied by
either a light chargino or a light smuon, setting interesting targets
for future collider searches.
\end{abstract}

%\pacs{}

\def\thefootnote{\arabic{footnote}}
\setcounter{page}{0}
\setcounter{footnote}{0}

\newpage

%%%%%%%%%%%%%%%%%%%%%%%%%%%%%%%%%%%%%%%%%%%%%%%%%%%%%%%%%%%%%%%%%%%%%%%%%%%%%%%
%%%%%%%%%%%%%%%%%%%%%%%%%%%%%%%%%%%%%%%%%%%%%%%%%%%%%%%%%%%%%%%%%%%%%%%%%%%%%%%

\section{Introduction}
\label{sec:intro}

Recently the ``MUON G-2'' collaboration~\cite{Grange:2015fou}
published the results of their Run~1 data~\cite{Abi:2021gix}
of the anomalous magnetic moment of the muon, $\amu := \edz \gmin2$,
which is within
$\fnalbnlsig\,\sig$ in agreement with  the older BNL result on \gmin2.
The combination of the two results yields
\begin{align}
\amu^{\rm exp} &= (11 659 \newcen \pm \newunc) \times 10^{-10}~.
\label{gmt-exp}
\end{align}
The Standard Model (SM) prediction of \amu\ is given by~\cite{Aoyama:2020ynm}
(based on \citeres{Aoyama:2012wk,Aoyama:2019ryr,Czarnecki:2002nt,Gnendiger:2013pva,Davier:2017zfy,Keshavarzi:2018mgv,Colangelo:2018mtw,Hoferichter:2019mqg,Davier:2019can,Keshavarzi:2019abf,Kurz:2014wya,Melnikov:2003xd,Masjuan:2017tvw,Colangelo:2017fiz,Hoferichter:2018kwz,Gerardin:2019vio,Bijnens:2019ghy,Colangelo:2019uex,Blum:2019ugy,Colangelo:2014qya}),
\begin{align}
\amu^{\rm SM} &= (11 659 181.0 \pm 4.3) \times 10^{-10}~.
\label{gmt-sm}
\end{align}
Accordingly, the discrepancy between the experimental value and the SM
prediction amounts to
\begin{align}
\Delta\amu \equiv \amu^{\rm exp} - \amu^{\rm SM}
&= (\newdiff \pm \newdiffunc) \times 10^{-10}~,
\label{gmt-diff}
\end{align}
corresponding to a $\newdiffsig\,\sig$ discrepancy.
At the time of the announcement of the MUON G-2 result, also a new
lattice calculation for the leading order hadronic
vacuum polarization (LO HVP) contribution to
$\amu^{\rm SM}$~\cite{Borsanyi:2020mff} has been published. This result,
however, had not been used in the new theory world
average, \refeq{gmt-sm}~\cite{Aoyama:2020ynm}.
In our analysis below we use
$\Delta\amu$ as given in \refeq{gmt-diff}.
The comparison of the new lattice result with previous results in the literature
and the assessment of the uncertainty estimate used in
\citere{Borsanyi:2020mff} are still a matter of debate, see e.g.\
\citeres{Lehner:2020crt,deRafael:2020uif}.
Possible implications
of the new lattice result
for the way physics beyond the SM (BSM) could manifest itself were discussed in
\citeres{Crivellin:2020zul,Keshavarzi:2020bfy}.
The impact of the new result for $\amu^{\rm exp}$
on possible scenarios of BSM physics that address its deviation
from the SM prediction will clearly depend on
how the theoretical prediction within the SM will settle during the next years.
If a discrepancy at the level of \refeq{gmt-diff} (or even stronger) will be
confirmed by future theoretical and experimental analyses,
$\amu$ will be a key observable for narrowing down the possible nature of BSM
physics.

Among the BSM theories under consideration
for accommodating a deviation at the level of \refeq{gmt-diff},
the Minimal Supersymmetric extension of the Standard Model
(MSSM)~\cite{Ni1984,Ba1988,HaK85,GuH86} is one of the most prominent candidates.
Supersymmetry (SUSY) predicts the existence of two scalar partners
for each SM fermion as well as fermionic partners for all SM bosons.
Contrary to the case of the SM, the theoretical structure of the
MSSM requires two Higgs doublets.
This results in five physical Higgs bosons instead of the single Higgs
boson in the SM. In the $\cp$-conserving case, considered in this
article, these are the light and heavy $\cp$-even Higgs bosons,
$h$ and $H$, the $\cp$-odd Higgs boson, $A$, and the charged Higgs
bosons,~$H^\pm$.
The neutral SUSY partners of the neutral Higgs and electroweak (EW) gauge
bosons give rise to the four neutralinos,~$\neu{1,2,3,4}$.  The corresponding
charged SUSY partners are the charginos,~$\cha{1,2}$.
The SUSY partners of the SM leptons and quarks are the scalar leptons
and quarks (sleptons, squarks), respectively.

In \citeres{CHS1,CHS2,CHS3,CHS4} some of us performed
an analysis of the EW sector of the MSSM,
taking into account all relevant experimental
data, i.e.\ data that is directly connected to the EW sector.
It was assumed that the Lightest
SUSY Particle (LSP)  is the lightest neutralino,~$\neu1$, with the
requirement that it is in agreement with the bounds on the Dark Matter (DM) content of the
universe~\cite{Go1983,ElHaNaOlSr1984}.
The experimental results employed in the analyses of
\citeres{CHS1,CHS2,CHS3,CHS4} comprise the direct searches at the
LHC~\cite{ATLAS-SUSY,CMS-SUSY}, the DM relic abundance~\cite{Planck},
the DM direct detection
experiments~\cite{XENON,LUX,PANDAX}
together with the deviation on the value
of the anomalous magnetic moment of the muon.%
\footnote{
In \citeres{CHS1,CHS2} the previous experimental result based on
\citeres{Keshavarzi:2019abf,Davier:2019can} was used,
while \citere{CHS1} was updated with the new world average on \gmin2\
in \citere{CHS3}.
}

In \citeres{CHS1,CHS2,CHS3,CHS4} five different scenarios were
analyzed, classified by the mechanism that has the main impact on the resulting
LSP relic density.
The scenarios differ by the nature of the Next-to-LSP
(NLSP). They comprise $\cha1$-coannhiliation, $\Slpm$-coannihilation with either
``left-'' or ``right-handed'' sleptons close in mass to the
LSP (``case-L'' and ``case-R'', respectively), wino DM, as well as
higgsino DM.
In the first
three scenarios the full amount of DM can be provided by the MSSM,
whereas in the latter two cases the measured DM density serves as an
upper limit. Requiring \refeq{gmt-diff} at the $2\,\sig$ level,
together with the collider and DM constraints,
results in upper limits on the LSP masses at the level of $\sim 500 \gev$
to $\sim 600 \gev$ for all five scenarios.
Corresponding upper limits on the mass of the NLSP are obtained for
only slightly higher mass values.

\smallskip
It is interesting to note that there is another
EW high-precision observable that shows a (slight) discrepancy between the experimental
result and the SM prediction, namely the mass of the $W$~boson, $\MW$. The
experimental world average is~\cite{PDG2020}
\begin{align}
\MWexp = 80.379 \pm 0.012 \gev~,
\label{mwexp}
\end{align}
whereas the SM predicts a value of
\begin{align}
\MWSM = 80.353 \pm 0.004 \gev~.
\label{mwsm}
\end{align}
For the central value of $\MWSM$ we use the implementation in the code
{\tt FeynHiggs}~\cite{feynhiggs}
(see below for details), while the quoted theoretical
uncertainty is based on an estimate of unknown higher-order
corrections~\cite{Awramik:2003rn} (in a comparison of the result in
the here used on-shell scheme with a result in the \MSbar\ scheme a difference of $6 \mev$ was
reported~\cite{Degrassi:2014sxa}). It is expected that this uncertainty can be reduced to
$\sim 0.001 \gev$ within the next decades, see \citere{Freitas:2019bre,Heinemeyer:2021rgq} and references therein.

Concerning the MSSM,
in \citeres{Heinemeyer:2006px,Heinemeyer:2007bw,Heinemeyer:2013dia} it
was shown that EW SUSY particles alone, provided that they are
sufficiently light, can induce significant shifts in the $\MWMSSM$ prediction
w.r.t.\ $\MWSM$  (taking into account the then
valid lower limits on the EW SUSY masses).
Those results motivate a combined analysis investigating whether the
discrepancy
in $\amu$ as given in \refeq{gmt-diff} and the difference between the current
value of $\MWexp$ as given in \refeq{mwexp} and the SM prediction of
\refeq{mwsm} could arise from loop corrections of the same type of SUSY
particles. While many papers
interpreted the observed \gmin2\ discrepancy of $\newdiffsig\,\sig$
in SUSY
models~\cite{CHS1,CHS2,CHS3,CHS4,Endo:2021zal,Iwamoto:2021aaf,Gu:2021mjd,VanBeekveld:2021tgn,Yin:2021mls,Wang:2021bcx,Abdughani:2021pdc,Cao:2021tuh,Ibe:2021cvf,Cox:2021gqq,Han:2021ify,Heinemeyer:2021zpc,Baum:2021qzx,Zhang:2021gun,Ahmed:2021htr,Athron:2021iuf,Aboubrahim:2021rwz,Chakraborti:2021bmv,Baer:2021aax,Altmannshofer:2021hfu,Chakraborti:2021squ,Zheng:2021wnu,Jeong:2021qey,Li:2021pnt,Dev:2021zty,Kim:2021suj,Ellis:2021zmg,Zhao:2021eaa,Frank:2021nkq,Shafi:2021jcg,Li:2021koa,Aranda:2021eyn,Aboubrahim:2021ily,Nakai:2021mha,Li:2021cte,Li:2021xmw,Lamborn:2021snt,Fischer:2021sqw,Forster:2021vyz,Ke:2021kgy,Ellis:2021vpp,Athron:2021dzk,Chakraborti:2021ynm,Chapman:2021gun,Aboubrahim:2021myl,Antoniadis:2021mqz,Acuna:2021rbg,Ali:2021kxa,Djouadi:2021wvb,Wang:2021lwi,Wang:2022rfd,Chakraborti:2022sbj,Boussejra:2022heb,Dermisek:2022hgh,Cao:2022chy,Gomez:2022qrb,Ahmed:2022ude,Chatterjee:2022pxf,Chakraborti:2022vds,Agashe:2022uih,Athron:2022uzz,Endo:2022qnm,Chigusa:2022xpq},
none of those papers analyzed $\MW$ in this context.

\smallskip
In the present paper we analyze the prediction of $\MWMSSM$ in view of the new
world average of $\amu^{\rm exp}$. We focus on the five MSSM
scenarios presented in \citeres{CHS1,CHS2}, well motivated by DM
constraints and characterized by relatively light EW SUSY particles and
a heavy colored supersymmetric spectrum.
Since the squarks and the gluino are assumed to be heavy in those scenarios,
in agreement with the experimental bounds from direct searches,
they yield
a negligible contribution
to $\MWMSSM$. All analyzed parameter points are in agreement with
the experimental result on the relic DM density,
which is imposed as an upper bound
(for the scenarios with $\cha1$-coannhiliation and $\Slpm$-coannihilation for
case-L and case-R this bound can be saturated, while for the scenarios with
wino DM and higgsino DM the obtained relic density stays below the measured
value), with DM direct detection bounds, and with LHC searches for light
EW SUSY particles.
In these scenarios we investigate the prediction for $\MWMSSM$
in combination with the contribution to $\De\amumssm$ w.r.t.\ the new
value for the discrepancy $\De\amu$ as given in \refeq{gmt-diff}.
We assess how well the MSSM prediction agrees
with the current experimental value for $\MWexp$. We analyze the
dependence on the relevant masses and parameters in the MSSM. Finally
we briefly discuss the impact of potential precision measurements
of $\MWexp$ at future $e^+e^-$ colliders such as the ILC, FCC-ee
or CEPC.

The paper is organized as follows.
In \refse{sec:model-constraints} we give a brief description of the
sectors of the MSSM relevant for our analysis. We review the
experimental constraints
applied to the MSSM parameter space and give a description of our
calculation of $\MWMSSM$. In the last part of this section we briefly
review the five scenarios that are analyzed in this paper. The numerical
results for the $\MWMSSM$ prediction in our scenarios, together with an
analysis of the relevant parameter dependences is given
in \refse{sec:results}. We conclude in \refse{sec:conclusion}.

%%%%%%%%%%%%%%%%%%%%%%%%%%%%%%%%%%%%%%%%%%%%%%%%%%%%%%%%%%%%%%%%%%%%
%%%%%%%%%%%%%%%%%%%%%%%%%%%%%%%%%%%%%%%%%%%%%%%%%%%%%%%%%%%%%%%%%%%%

\section {The model, experimental constraints and \boldmath{$\MW$}}
\label{sec:model-constraints}

\subsection{The model}
\label{sec:model}

A detailed description of our conventions used for
the EW sector of the MSSM can be found in \citere{CHS1}. Here we just list the
input parameters and masses that are relevant for our analysis. Throughout this
paper we assume that all parameters are real, i.e.\ we
do not incorporate possible $\CP$-violating effects that can be induced
via SUSY loop corrections with complex parameters.

The masses and mixings of the charginos and neutralinos are determined
by $U(1)_Y$ and $SU(2)_L$ gaugino soft SUSY-breaking mass parameters
$M_1$ and $M_2$, the Higgs/higgsino mass parameter $\mu$, and $\tb$, the
ratio of the
vacuum expectation values (vevs) of the two Higgs doublets of the MSSM,
$\tb \equiv v_2/v_1$.
The four neutralino masses are ordered as $\mneu1 < \mneu2 < \mneu3 <\mneu4$.
Similarly, the two chargino-masses are denoted as $\mcha1 < \mcha2$.
As explained in \citere{CHS1}, it is sufficient for our analysis to focus
on positive values for $M_1$, $M_2$ and $\mu$.

For the sleptons, as in \citere{CHS1}, we choose common soft
SUSY-breaking parameters for all three generations, $\mL$ and $\mR$.
We take the trilinear coupling
$A_l$ ($l = e, \mu, \tau$) to be zero.
We follow the convention that $\Sl_1$ ($\Sl_2$) has the
large ``left-handed'' (``right-handed'') component.
Besides the symbols that refer to equal values
for all three generations, $\msl1$ and $\msl2$,
we also explicitly use the scalar electron, muon and tau masses,
$\mse{1,2}$, $\msmu{1,2}$ and $\mstau{1,2}$.

We assume that the colored sector of the MSSM is
significantly heavier than the EW sector, and does not play a role in this analysis since it
yields a negligible contribution to $\amumssm$ and $\MWMSSM$.
Since the colored particles are assumed to be very heavy, they are not
affected by the LHC limits from direct searches~\cite{ATLAS-SUSY,CMS-SUSY}.
In particular we have chosen the scalar top/bottom sector
parameters in such a way that the radiative corrections to the
mass of the light
$\cp$-even Higgs boson yield a value in agreement with the experimental
data, $\Mh \sim 125 \gev$ (taking into account a theory
uncertainties of $\pm 3 \gev$, which as a conservative estimate amounts to
about twice the value obtained in \citere{mh-unc}).
The determination of the stop/sbottom MSSM parameters has been performed
with {\tt FeynHiggs-2.18.0}~\cite{feynhiggs,mhiggslong,mhiggsAEC,Mh-logresum,mh-unc,feynhiggs-new}.
The resulting  stop masses are found to be heavier than
$2\tev$~\cite{Bagnaschi:2017tru,Slavich:2020zjv},
in agreement with the above assumption of a heavy colored sector.

The mass of the $\cp$-odd Higgs boson, $\MA$, has been
assumed to be heavy ($\gsim 2\tev$) which ensures its compatibility
with the experimental bounds from the LHC. As a consequence, effects
from $H$-/$A$-pole
annihilation of DM in the early universe are absent in our analysis.

%%%%%%%%%%%%%%%%%%%%%%%%%%%%%%%%%%%%%%%%%%%%%%%%%%%%%%%%%%%%%%%%%%%%%%%%%%

\subsection {Relevant constraints}
\label{sec:constraints}

Concerning the experimental constraints taken into account, we
follow \citere{CHS1}. These comprise

\begin{itemize}

\item Vacuum stability constraints:\\
All points are checked to possess an EW vacuum that has the correct
properties and is stable,
e.~g.\ avoiding charge and color breaking minima. This check is performed with
the public code {\tt Evade}~\cite{Hollik:2018wrr,Ferreira:2019iqb}.

\item Constraints from the LHC and LEP:\\
EW SUSY searches at the LHC are taken into account
as described in \citere{CHS1}. This has mostly been done via
\CM~\cite{Drees:2013wra,Kim:2015wza, Dercks:2016npn}, where many
analyses were newly implemented~\cite{CHS1}.
In this context also the code {\tt SDECAY}~\cite{Muhlleitner:2003vg} as
in included in {\tt SUSYHIT-1.5a} has been used.
The points are furthermore required to satisfy the $\chapm1$ mass limit
from LEP~\cite{lepsusy}.

\item
Dark matter relic density constraints:\\
We use the latest result from Planck~\cite{Planck} as an upper limit.
The relic density in the MSSM is evaluated with
\MO{\texttt{-5.0.8}}~\cite{Belanger:2001fz,Belanger:2006is,Belanger:2007zz,Belanger:2013oya}.

\item
Dark matter direct detection constraints:\\
We employ the constraint on the spin-independent
DM scattering cross-section $\ssi$ from the
XENON1T~\cite{XENON} experiment,
evaluating the theoretical prediction for $\ssi$ using
\MO. A combination with other direct detection experiments would yield
only very slightly
stronger limits, with a negligible impact on our results.

\item
We indicate in the plots the new result for $\De\amu$ as given in
\refeq{gmt-diff}, which is applied at the $\pm 2\sig$ level
as a constraint that the parameter points have to pass.
The evaluation has been done with the code
{\tt GM2Calc-1.7.5}~\cite{Athron:2015rva}, incorporating two-loop corrections
from \citeres{vonWeitershausen:2010zr,Fargnoli:2013zia,Bach:2015doa}
(see also \citeres{Heinemeyer:2003dq,Heinemeyer:2004yq}).

\end{itemize}

In addition, we have verified using
{\tt HiggsSignals-2.6.1}~\cite{Bechtle:2013xfa,Bechtle:2014ewa,Bechtle:2020uwn}
that the properties of the Higgs state at $\sim 125$ GeV, which could be
modified by presence of light SUSY EW states, are in agreement with the
current measurements.

%%%%%%%%%%%%%%%%%%%%%%%%%%%%%%%%%%%%%%%%%%%%%%%%%%%%%%%%%%%%%%%%%%%%%%%%%%

\subsection{The \boldmath{$W$}-boson mass}
\label{sec:mw}

The mass of the $W$~boson can be predicted from muon decay, which relates
$\MW$ to three extremely precisely measured quantities, namely the Fermi
constant, $G_\mu$, the fine structure constant, $\al$, and the mass of the
$Z$~boson, $\MZ$. Within the SM and many extensions of it, in particular the
MSSM, this relation can be used to predict $\MW$ via the expression%
\footnote{See e.g.\ \citere{Diessner:2019ebm} for the case of a model where the
lowest-order prediction for $\MW$ is modified.}
\begin{align}
\MW^2 = \MZ^2 \LV \frac{1}{2} +
\sqrt{\frac{1}{4} - \frac{\pi\,\al}{\sqrt{2}\,G_\mu\,\MZ^2}
\LB 1 + \De r(\MW, \MZ, \mt, \ldots) \RB } \RV~,
\label{eq:mwpred}
\end{align}
where the quantity $\De r$ is zero at lowest order.
It comprises loop corrections to muon decay in the considered model, where the
ellipsis in \refeq{eq:mwpred} denotes the specific particle content of the
model. Since $\De r$ is a function of $\MW$ itself, it is convenient to evaluate
\refeq{eq:mwpred} via an iterative procedure.

The SM prediction for $\De r$ includes contributions at the complete
one-loop~\cite{Sirlin:1980nh,Marciano:1980pb} and the complete
two-loop level~\cite{Djouadi:1987gn,Djouadi:1987di,Kniehl:1989yc,Halzen:1990je,Kniehl:1991gu,Kniehl:1992dx,Halzen:1991ik,Freitas:2000gg,Freitas:2002ja,Awramik:2002wn,Awramik:2003ee,Onishchenko:2002ve,Awramik:2002vu,Bauberger:1996ix,Bauberger:1997ey,Awramik:2006uz},
as well as partial higher-order corrections up to four-loop
order~\cite{Avdeev:1994db,Chetyrkin:1995ix,Chetyrkin:1995js,Chetyrkin:1996cf,Faisst:2003px,vanderBij:2000cg,Boughezal:2004ef,Schroder:2005db,Chetyrkin:2006bj,Boughezal:2006xk}.%
\footnote{See also \citeres{Weiglein:1998jz,AchimDipl,Chen:2020xzx,Chen:2020xot}
for further higher-order contributions
involving fermion loops.}
Our prediction for $\MW$ in the MSSM is based on the full one-loop result for
$\De r$~\cite{Heinemeyer:2006px,Heinemeyer:2007bw,Heinemeyer:2013dia}
(see also \citere{Chankowski:1993eu}),
supplemented by the leading two-loop
corrections~\cite{Djouadi:1996pa,Djouadi:1998sq,Haestier:2005ja}.
The leading one- and two-loop contributions arise from isospin splitting
between different SUSY particles and enter via the quantity $\De\rho$, which
receives contributions from the $W$-boson and $Z$-boson self-energies at
vanishing external momentum.
At the one-loop level the squarks enter only via self-energy contributions,
i.e.\ predominantly via $\De\rho$. The same is true for the corresponding
contribution of pure slepton loops, while the contributions of the chargino and
neutralino sector enter also via vertex and box diagrams. In our MSSM prediction
for $\MW$ the contributions involving SUSY particles are combined with all
available SM-type contributions up to the four-loop level as described above.
This ensures that the
state-of-the-art SM prediction is recovered in the decoupling limit where all
SUSY mass scales are heavy.
For our analysis we use the implementation of the SM and SUSY contributions
in the code {\tt FeynHiggs} as described in \citere{Stal:2015zca}.

Concerning the prediction for $\MW$ in the SM,
the central value of $\MWSM$ quoted in \refeq{mwsm}
(which agrees with the result obtained
from the fit formula in \citere{Awramik:2003rn} within the theoretical
uncertainty)
has been obtained
for the following input parameters:
\begin{align}
% &\alpha(0)^{-1} &=&~137.035999084,     \quad \hspace{0.2cm} M_Z
% &=&~91.1876~\mathrm{GeV}, \quad \hspace{1cm} G_F &=&~1.166378 \cdot
% 10^{-5}~~\mathrm{GeV}^{-2},  \nonumber \\
% &\Delta \alpha^{(5)}_{\mathrm{had}} (M_Z)
% &=&~0.02766, \quad \hspace{1.5cm} \mt
% &=&~172.76~\mathrm{GeV}, \quad  \hspace{1cm} \MH^{\mathrm{SM}}
% &=&~125.09 \gev,
%\label{eq:inputs}\\
%&\Delta \alpha_{\mathrm{lept}}  &=&~0.031497687, \quad  m_b(m_b)
%&=&~4.18~\mathrm{GeV}, \quad \hspace{1cm} \alpha_s(\MZ) &=&~0.1179 .
%\nonumber
 \alpha(0)^{-1} &= 137.035999084 \,,
 &\MZ &= 91.1876 \gev \,,
 &G_F &= 1.166378 \cdot 10^{-5} \gev^{-2} \,,  \nonumber \\
 \De \al^{(5)}_{\mathrm{had}} (\MZ)  &= 0.02766 \,,
 &\mt &= 172.76 \gev \,,
 &\MH^{\mathrm{SM}} &= 125.09 \gev \,,
\label{eq:inputs}\\
 \De \al_{\mathrm{lept}} &= 0.031497687 \,,
 &m_b(m_b) &= 4.18 \gev\,,
 &\al_s(\MZ) &= 0.1179 \,.
\nonumber
\end{align}
All these values are from Ref.~\cite{ParticleDataGroup:2020ssz}, except for
$\Delta \alpha_{\mathrm{lept}}$, which is taken from
Ref.~\cite{Steinhauser:1998rq}.
Besides the ``intrinsic'' theoretical uncertainties from unknown
higher-order corrections,
see the estimate in \refeq{mwsm} for the SM, the predictions in the SM and the MSSM are
also affected by theoretical uncertainties that are induced by the experimental
errors of the input parameters.
For the SM case the latter ``parametric'' uncertainties
can be estimated by recomputing $\MWSM$ for numerical values of the input parameters
that are shifted with respect to the ones quoted in \refeq{eq:inputs} by their
experimental errors (where we use the values given in
\citere{ParticleDataGroup:2020ssz} unless
otherwise specified): for the case of the top-quark mass, $\mt$,
a variation of $\pm 1 \gev$~\footnote{
This value is indicated for illustration. The parametric uncertainty of the
top-quark mass has to be assessed on the basis of the experimental error
of the measured mass parameter of $\pm 0.30\gev$
at the $1\,\sigma$ level~\cite{ParticleDataGroup:2020ssz}
in combination with the systematic uncertainty that
is associated with relating the measured quantity to a theoretically
well-defined top-quark mass.}
changes $\MWSM$ by $\pm$ 6 MeV; a shift of $\pm 0.0010$ from the central
value $\alpha_s(M_Z)$ yields a variation of $\pm 0.7$ MeV in $\MWSM$;
the uncertainty on $M_Z$ is of $\pm 0.0021$ GeV and its impact on
$\MWSM$ is of $\simeq \pm 2.7$ MeV; the uncertainty of $0.00007$ on the
value of $\Delta \alpha^{(5)}_{\mathrm{had}} (M_Z)$ yields a variation
of $\simeq 1.2$ MeV; finally, varying the Higgs mass by 1~GeV results in
a shift of $\simeq 0.4$ MeV.

Concerning the theoretical uncertainties
of the MSSM prediction for $\MW$, our implementation described above
is such that in the decoupling limit
the intrinsic and
parametric theoretical uncertainties of the $\MW$ prediction are the same as for
the SM case. If some of the SUSY particles are relatively light the intrinsic
theoretical uncertainties can be somewhat larger, depending on the mass scales
of the SUSY parameters, see, e.g., the discussion in
\citere{Heinemeyer:2006px}. The sensitivity of the $\MW$ prediction to
variations of the SUSY parameters is usually not discussed as a
parametric uncertainty, but as
an indication of the sensitivity of a precise measurement of $\MW$ for
constraining the allowed range of SUSY parameters.
In our numerical analysis below the intrinsic theoretical uncertainty of the
prediction for $\MW$ in the MSSM from unknown higher-order corrections and the
parametric uncertainty from varying the SM input parameters will not be
displayed.
On the other hand, we also show for comparison the SM prediction, where the
current intrinsic theoretical uncertainty from unknown higher-order
corrections of
$\pm 4 \mev$, see \refeq{mwsm}, is indicated as a horizontal band.

We furthermore indicate in our plots below the
current experimental value for $\MW$ and its
$\pm 1 \sig$ band as given in \refeq{mwexp}.
For illustration also a projected
future uncertainty that can be expected at the
ILC of $\de\MW^{\rm ILC} = 0.003 \gev$~\cite{Fujii:2019zll}
is shown.
In principle an even
higher precision
of \order{1\mev} could be achieved
at the
FCC-ee~\cite{Gomez-Ceballos:2013zzn,Abada:2019zxq} or at the
CEPC~\cite{CEPCStudyGroup:2018ghi,CEPCPhysics-DetectorStudyGroup:2019wir},
but in the projections that were carried out
theory uncertainties affecting the measurement were not taken into account.
Those theoretical uncertainties may give rise to a systematic experimental
uncertainty that dominates over the expected statistical
uncertainty~\cite{Freitas:2019bre,Heinemeyer:2021rgq}.

%%%%%%%%%%%%%%%%%%%%%%%%%%%%%%%%%%%%%%%%%%%%%%%%%%%%%%%%%%%%%%%%%%%%%%%%%%

\subsection{Parameter scan}
\label{sec:scan}

We use the parameter sets that were obtained in \citeres{CHS1,CHS2,CHS3}
from a scan of the EW MSSM parameter space making use of the code
{\tt SuSpect-2.43}~\cite{Djouadi:2002ze}.
The applied constraints discussed above, in particular the compatibility
with $\De\amu$ as given in \refeq{gmt-diff} at the $2\,\sigma$ level
and with the upper bound on the DM relic density, give rise to
lower and upper limits on the relevant neutralino, chargino and slepton masses.
As detailed in \citeres{CHS1,CHS2} five scan regions cover the relevant
parameter space:

\begin{description}
\item
{\bf (A) Mixed bino/wino DM with \boldmath{$\chapm1$}-coannihilation}\\
\begin{align}
  100 \gev \leq M_1 \leq 1 \tev \;,
  \quad M_1 \leq M_2 \leq 1.1 M_1\;, \notag \\
  \quad 1.1 M_1 \leq \mu \leq 10 M_1, \;
  \quad 5 \leq \tb \leq 60, \; \notag\\
  \quad 100 \gev \leq \mL \leq 1 \tev, \; %\notag\\
  \quad \mR = \mL~.
\label{cha-coann}
\end{align}

\item
{{\bf Bino DM with \boldmath{$\Slpm$}-coannihilation region}}\\
%  \small
{\bf (B)} Case-L: SU(2) doublet
\begin{align}
  100 \gev \leq M_1 \leq 1 \tev \;,
  \quad M_1 \leq M_2 \leq 10 M_1 \;, \notag\\
  \quad 1.1 M_1 \leq \mu \leq 10 M_1, \;
  \quad 5 \leq \tb \leq 60, \; \notag\\
  \quad M_1 \leq \mL \leq 1.2 M_1, %\notag\\
  \quad M_1 \leq \mR \leq 10 M_1~.
\label{slep-coann-doublet}
\end{align}

{\bf (C)} Case-R: SU(2) singlet
\begin{align}
  100 \gev \leq M_1 \leq 1 \tev \;,
  \quad M_1 \leq M_2 \leq 10 M_1 \;, \notag \\
  \quad 1.1 M_1 \leq \mu \leq 10 M_1, \;
  \quad 5 \leq \tb \leq 60, \; \notag\\
  \quad M_1 \leq \mR \leq 1.2 M_1,\; %\notag\\
  \quad M_1 \leq \mL \leq 10 M_1~.
\label{slep-coann-singlet}
\end{align}

\item
{\bf (D) Higgsino DM}\\
\begin{align}
  100 \gev \leq \mu \leq 1.2 \tev \;,
  \quad 1.1 \mu \leq M_1 \leq 10 \mu\;, \notag \\
  \quad 1.1  \mu \leq M_2 \leq 10 \mu, \;
  \quad 5 \leq \tb \leq 60, \; \notag\\
  \quad 100 \gev \leq \mL, \mR \leq 2  \tev~.
\label{higgsino-dm}
\end{align}

\item
{\bf (E) Wino DM}\\
\begin{align}
  100 \gev \leq M_2 \leq 1.5 \tev \;,
  \quad 1.1 M_2 \leq M_1 \leq 10 M_2\;, \notag \\
  \quad 1.1 M_2 \leq \mu \leq 10 M_2, \;
  \quad 5 \leq \tb \leq 60, \; \notag\\
  \quad 100 \gev \leq \mL, \mR \leq 2 \tev~.
\label{wino-dm}
\end{align}
\end{description}

For each of the five scenarios a data sample of \order{10^7} points
was generated by scanning randomly over the input parameter
ranges specified above, using a flat prior for all parameters.
Since we have assumed the colored SUSY sector to be heavy, the
parameters of the squark and gluino sectors are not varied in the scan
(see above for the prescription that was used for obtaining a prediction
for the mass of the SM-like Higgs boson that is in agreement with the
experimental result).
We have checked for our scan points that the contribution
from colored SUSY particles to $\MWMSSM$ is indeed negligible.

%%%%%%%%%%%%%%%%%%%%%%%%%%%%%%%%%%%%%%%%%%%%%%%%%%%%%%%%%%%%%%%%%%%%%%%%%%

\section{Results}
\label{sec:results}

In the following we present the results for the scan points that are
allowed by the
experimental and theoretical constraints specified in \refse{sec:constraints}
in the five scenarios defined above. In particular, the displayed scan
points are in agreement with $\De\amu$, as given in \refeq{gmt-diff}, at
the~1 and $2\,\sig$ level.

%%%%%%%%%%%%%%%%%%%%%%%%%%%%%%%%%%%%%%%%%%%%%%%%%%%%%%%%%%%%%%%%%%%%%%%%%%%%%%%

%%%%%%%%%%%%%%%%%%%%%%%%%% F I G U R E %%%%%%%%%%%%%%%%%%%%%%%%%%%%%%%%%%%%%%%
\begin{figure}[htb!]
\begin{center}
\includegraphics[width=0.9\textwidth]{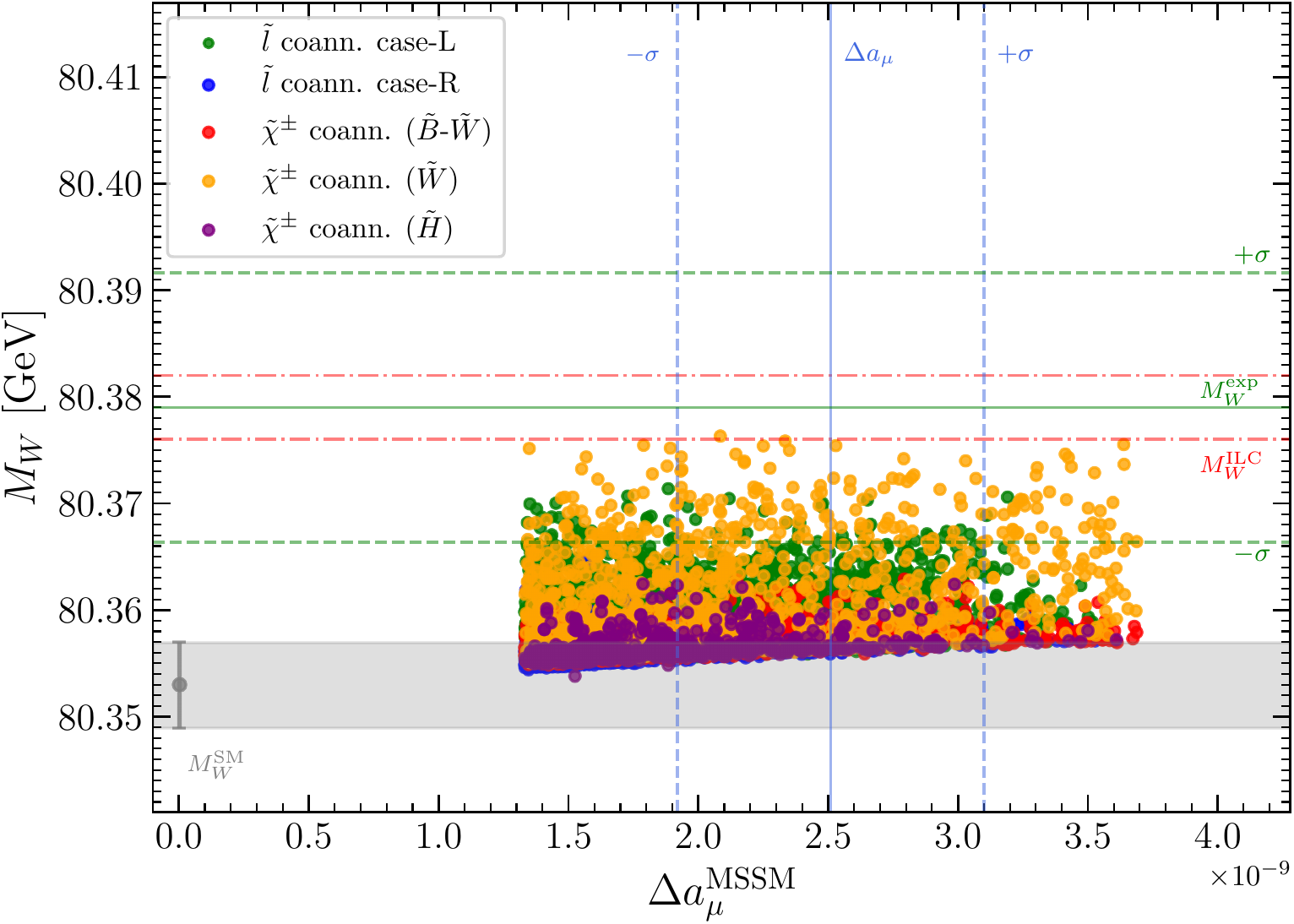}
\caption{\label{fig:amu-mw}
Results for the five considered scenarios in the $\De\amumssm$--$\MW$ plane,
where the prediction for $\De\amumssm$ has been evaluated with the code
{\tt GM2Calc-1.7.5}~\cite{Athron:2015rva}.
The points for the $\Slpm$-coannihilation case-L and case-R,
the $\cha1$-coannihilation case, the wino and the higgsino case are shown in
green, blue, red, orange and violet, respectively.
The vertical blue lines indicate the central value of $\De\amu$
as given in \refeq{gmt-diff} (solid) and its $\pm 1\,\sig$ range (dashed).
The displayed points are restricted to the $\pm 2\,\sig$ range of $\De\amu$.
The horizontal lines indicate the current central value for $\MWexp$
(solid green), the current $\pm 1\,\sig$ uncertainties (green dashed)
and the anticipated ILC $\pm 1\,\sig$ (red dot-dashed) uncertainties.
The SM prediction is shown as a point for $\De\amumssm = 0$, while the gray
band indicates the theoretical uncertainty of the SM prediction for $\MW$
from unknown higher-order corrections.
}
\end{center}
\vspace{-1em}
\end{figure}
%%%%%%%%%%%%%%%%%%%%%%%%%% F I G U R E %%%%%%%%%%%%%%%%%%%%%%%%%%%%%%%%%%%%%%%

In \reffi{fig:amu-mw} we show in the $\De\amumssm$--$\MW$ plane
the results for the five scenarios corresponding to the
$\Slpm$-coannihilation case-L and case-R,
the $\cha1$-coannihilation case, the wino and the higgsino case.
The prediction for $\De\amumssm$ has been evaluated with the code
{\tt GM2Calc-1.7.5}~\cite{Athron:2015rva}.
The vertical solid blue line indicates the value of
$\De\amu$ as given in \refeq{gmt-diff}, while its
$\pm 1\,\sig$ range is indicated by the blue dashed vertical lines.
The displayed points are restricted to the $\pm 2\,\sig$ range of $\De\amu$.
The horizontal lines indicate the current central value for $\MWexp$
(solid green), the current $\pm 1\,\sig$ uncertainties (green dashed)
and the anticipated ILC $\pm 1\,\sig$ (red shaded) uncertainties.
The SM prediction is shown in gray, including the theoretical uncertainty
from unknown higher-order corrections.

One can observe in all scenarios a lower limit on $\MWMSSM$
that for small $\De\amumssm$, corresponding to heavy EW SUSY masses,
recovers the SM
prediction (within $\sim 1 \mev$; this offset would be absent for even
smaller values
of $\De\amumssm$). The lower limit rises for increasing $\De\amumssm$ by
up to $\sim 3 \mev$.
Thus, the relatively light SUSY particles that are required for larger values
of $\De\amumssm$ give rise to a slight increase in the prediction for
$\MW$ that is independent of the variation of the other parameters in
the scan. While this lower limit on the predicted value of $\MW$ is very
similar in the five DM scenarios, there are important differences in
the highest $\MWMSSM$ values that are reached.
The largest predicted values of $\MWMSSM$, nearly reaching the current
central value of $\MWexp$,
are obtained for the wino DM case.
Accordingly, for the wino DM case the electroweak sector of the MSSM behaves
in such a way that the predicted values for $\MW$ and the anomalous magnetic
moment of the muon can simultaneously be very close to the present experimental
central values, while respecting all other constraints on the model.
For the $\Slpm$-coannihilation case-L the highest obtained $\MWMSSM$ values
are somewhat lower but still within the current $\pm 1\,\sig$ range of $\MWexp$.
On the other hand, for the other three scenarios we find significantly lower
predicted values of $\MW$.

These results open interesting perspectives for
discriminating between different DM scenarios with the
anticipated future accuracy on $\MW$.
Depending of course on the future central value of $\MW$,
the wino DM scenario could potentially be singled out as the only
scenario yielding a good agreement with the $\MW$
measurement. As we will discuss in more detail below, an $\MWMSSM$ prediction
near the current experimental central value implies that at least some of the
SUSY particles have to be relatively light, offering good prospects for
the SUSY searches at the LHC and future colliders.
If instead the future central value of $\MWexp$ turns out to be
substantially lower, outside of the current $\pm 1\,\sig$ range, all the
considered DM
scenarios could still be in agreement with the $\MW$ measurement.
On the other hand, a higher central value of $\MWexp$ would require
larger contributions from the EW SUSY states, possibly via further
relaxed assumptions on the EW SUSY parameters,
or additional contributions from the colored sector (as discussed above,
the latter possibility is not considered in the present study).

%%%%%%%%%%%%%%%%%%%%%%%%%% F I G U R E %%%%%%%%%%%%%%%%%%%%%%%%%%%%%%%%%%%%%%%
\begin{figure}[htb!]
\begin{center}
\includegraphics[width=0.9\textwidth]{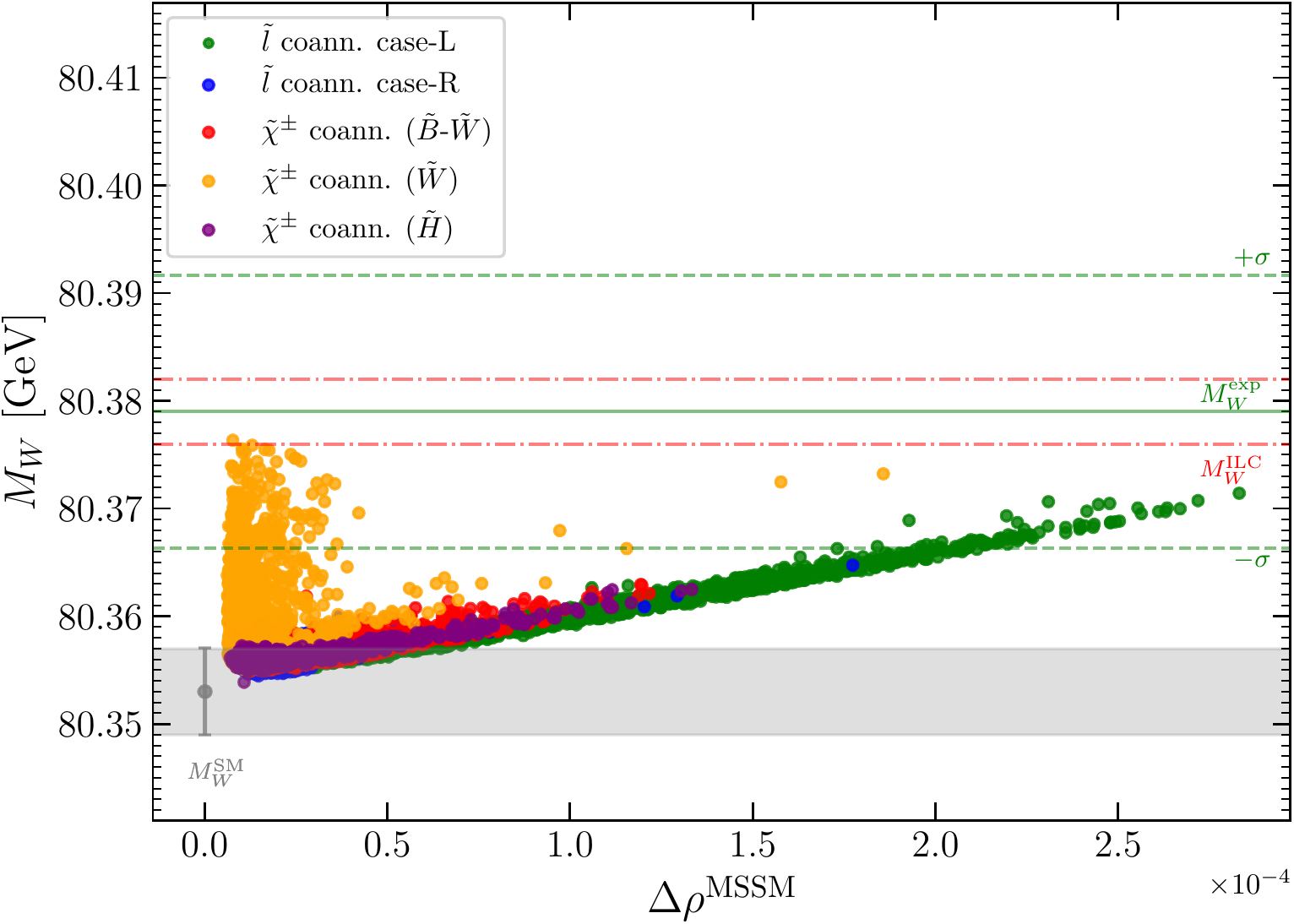}
\caption{\label{fig:deltarho-mw}
Results for the five scenarios in the $\De\rho^{\mathrm{MSSM}}$--$\MW$ plane.
The horizontal lines and the color coding are
as in \protect\reffi{fig:amu-mw}.
}
\end{center}
\end{figure}
%%%%%%%%%%%%%%%%%%%%%%%%%% F I G U R E %%%%%%%%%%%%%%%%%%%%%%%%%%%%%%%%%%%%%%%

In \reffi{fig:deltarho-mw} we analyze in more detail the origin of the
contributions to $\MWMSSM$. We show the $\MW$ prediction as a function
of $\De\rho^{\mathrm{MSSM}}$, which denotes the SUSY contributions to
the quantity $\De\rho$. As discussed above,
the contribution of pure slepton loops enters only via self-energies, and in
particular via the SUSY contributions to $\De\rho$. On the other
hand,
``mixed'' slepton/chargino/neutralino
contributions enter via the vertex and box corrections to $\De r$.
One can see in \reffi{fig:deltarho-mw} that the largest corrections to
$\MWMSSM$ that were found for wino DM and the $\Slpm$-coannihilation
case-L in \reffi{fig:amu-mw}
have different origins. Wino DM, which naturally has at least
one light neutralino and one light chargino in the spectrum, can yield
sizable contributions to $\MWMSSM$ even for small
$\De\rho^{\mathrm{MSSM}}$. This is caused by numerically important one-loop
vertex and box contributions.
On the other hand, for $\Slpm$-coannihilation, which naturally has light
sleptons in
the spectrum, large corrections to $\MWMSSM$ are correlated with large
contributions of
$\De\rho$, while the vertex and box contributions in this case are
sub-dominant.
We have explicitly verified these features by enabling and
disabling the respective contributions in the $\MWMSSM$ calculation.

Consequently, in the following we discuss the dependence of $\MWMSSM$
on the various SUSY particle masses. We will focus our discussion on the two
scenarios that show an appreciable contribution to $\MWMSSM$, wino DM
and the $\Slpm$-coannihilation case-L. For the other scenarios
the impact of an improved accuracy of the $\MW$ measurement
on the parameter space can be summarized as follows:
for an experimental central value that is close to
the SM prediction the discriminating power between different SUSY scenarios will
be limited; on the other hand, if the experimental central value stays close to
the current value those scenarios will be disfavored by the $\MW$ measurement
(unless relatively light colored particles would yield an upward shift in the
$\MW$ prediction).

%%%%%%%%%%%%%%%%%%%%%%%%%% F I G U R E %%%%%%%%%%%%%%%%%%%%%%%%%%%%%%%%%%%%%%%
\begin{figure}[htb!]
\begin{center}
\includegraphics[width=0.77\textwidth]{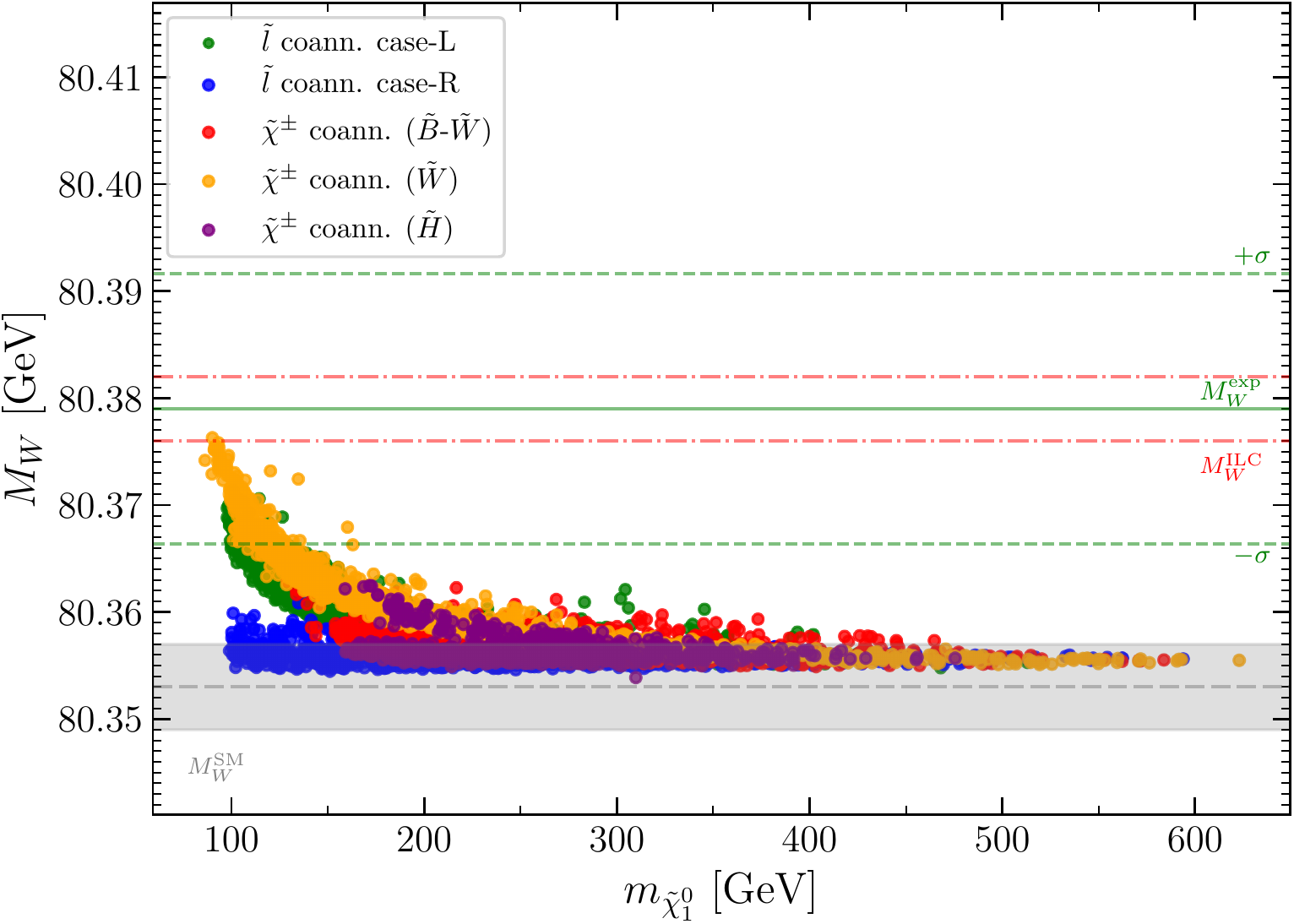}
\caption{\label{fig:mneu1-mw}
Results for the five scenarios in the $\mneu1$--$\MW$ plane.
The horizontal lines and the color coding are
as in \protect\reffi{fig:amu-mw}.
}
\end{center}
\end{figure}
%%%%%%%%%%%%%%%%%%%%%%%%%% F I G U R E %%%%%%%%%%%%%%%%%%%%%%%%%%%%%%%%%%%%%%%

%%%%%%%%%%%%%%%%%%%%%%%%%% F I G U R E %%%%%%%%%%%%%%%%%%%%%%%%%%%%%%%%%%%%%%%
\begin{figure}[htb!]
\begin{center}
\includegraphics[width=0.77\textwidth]{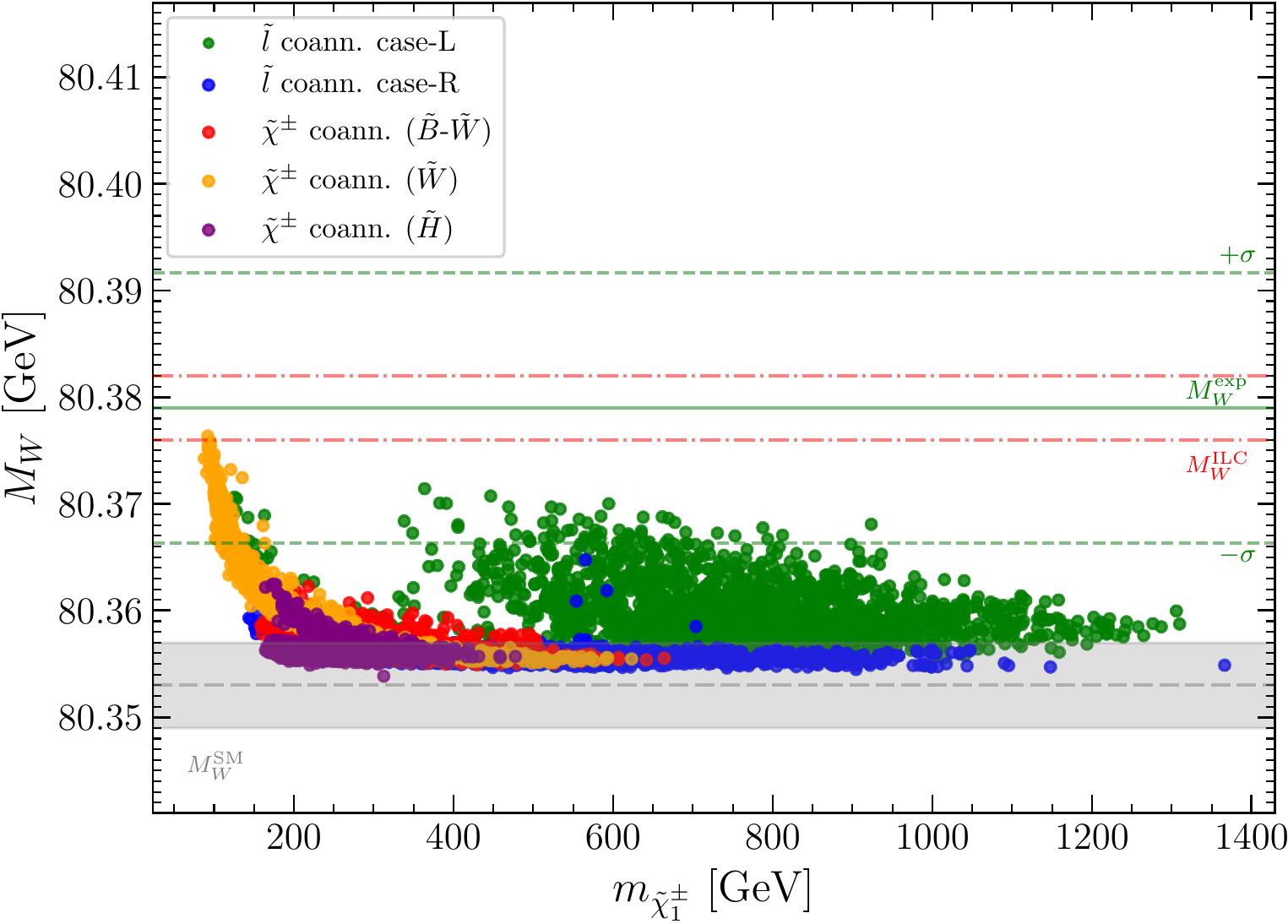}
\caption{\label{fig:mcha1-mw}
Results for the five scenarios in the $\mcha1$--$\MW$ plane.
The horizontal lines and the color coding are as in \protect\reffi{fig:amu-mw}.
}
\end{center}
\end{figure}
%%%%%%%%%%%%%%%%%%%%%%%%%% F I G U R E %%%%%%%%%%%%%%%%%%%%%%%%%%%%%%%%%%%%%%%

We start our discussion of the SUSY mass dependences of $\MWMSSM$
in \reffi{fig:mneu1-mw} with the prediction of the $W$-boson mass as a
function of $\mneu1$.
In our scan we find values of the LSP mass between $\sim 100 \gev$ and
up to $\sim 600 \gev$,
where the specific values depend on the scenario (see also the discussion
in~\citeres{CHS1,CHS2,CHS3,CHS4}). In particular, only for
$\Slpm$-coannihilation and wino DM very low values of $\mneu1$ are
realized. The large contributions to $\MWMSSM$ in the wino DM scenario
and for the $\Slpm$-coannihilation case-L are reached only for
$\mneu1 \lsim 150 \gev$. For larger LSP masses the MSSM prediction
for $\MW$ approaches the SM limit, indicating a decoupling effect of the
SUSY contributions. The low $\mneu1$ values that can
bring the $\MW$ prediction close to the experimental central value
provide an interesting target that can be probed via
SUSY searches at the LHC and future colliders.

In \reffi{fig:mcha1-mw} we show the dependence of $\MWMSSM$ on the
lightest chargino mass, $\mcha1$.
The largest  contributions to $\MWMSSM$ are obtained for the smallest
chargino masses, $\mcha1 \lsim 200 \gev$.
On the other hand, for the $\Slpm$-coannihilation case-L $\MW$ values
within the current $1\,\sig$ bound are reached for masses up to
$\mcha1 \lsim 800 \gev$. The visible ``hole'' in this scenario for
$200 \gev \lsim \mcha1 \lsim 350 \gev$, where no scan points passed the
applied constraints, is due to the experimental constraints from the
LHC. In particular, the strongest impact comes from the ATLAS
3l+\met~search~\cite{ATLAS:2019lff}, which is sensitive to the
production of a pair of $\tilde{\chi}_1^{\pm} \tilde{\chi}_2^0$ with
subsequent decay via sleptons.

\begin{figure}[htb!]
\begin{center}
\includegraphics[width=0.77\textwidth]{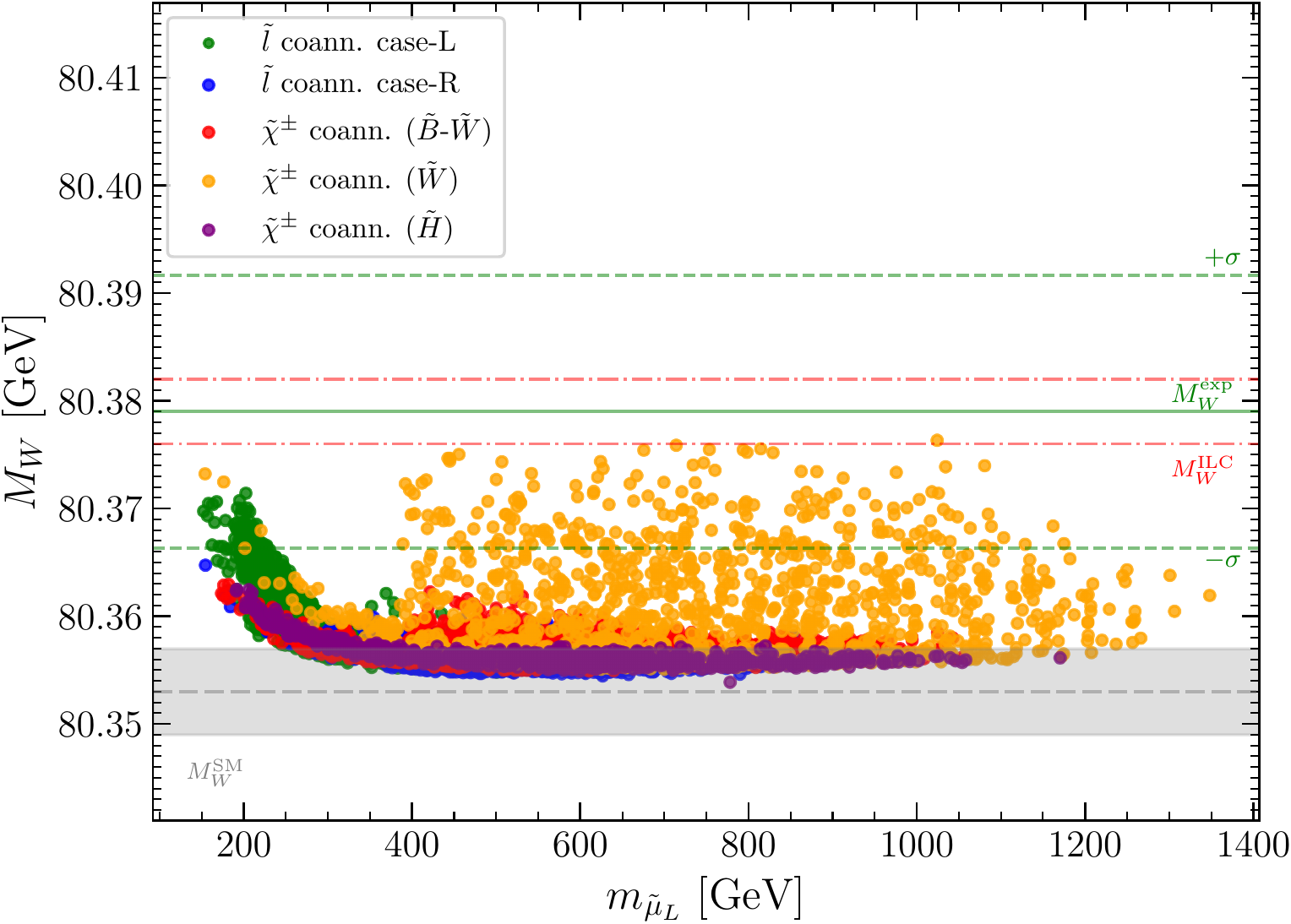}
\caption{\label{fig:msmuL-mw}
Results for the five scenarios in the $\msmu{L}$--$\MW$ plane.
The horizontal lines and the color coding are as in \protect\reffi{fig:amu-mw}.
}
\end{center}
\end{figure}

\begin{figure}[htb!]
\begin{center}
\includegraphics[width=0.77\textwidth]{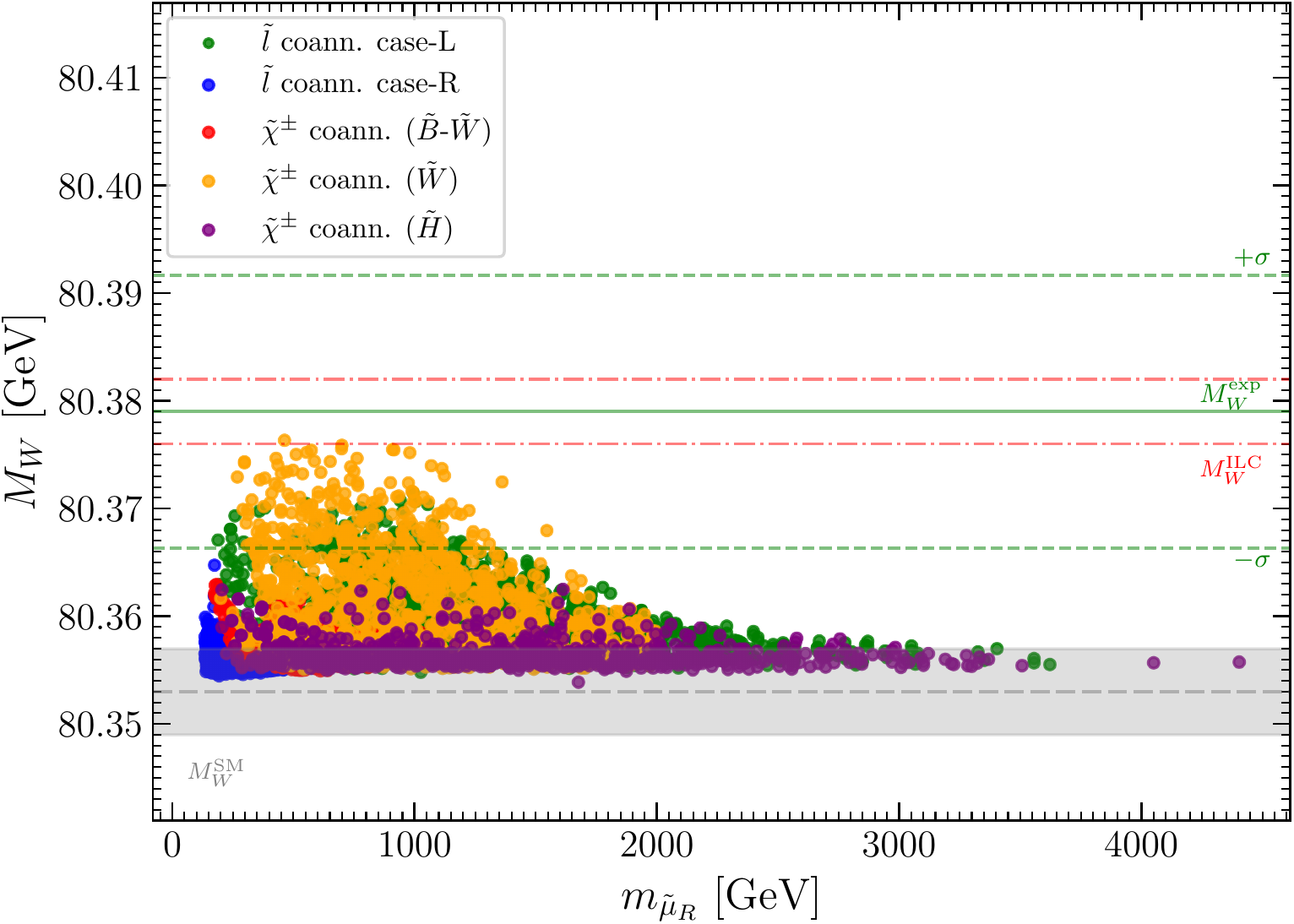}
\caption{\label{fig:msmuR-mw}
Results for the five scenarios in the $\msmu{R}$--$\MW$ plane.
The horizontal lines and the color coding are
as in \protect\reffi{fig:amu-mw}.
}
\end{center}
\end{figure}

Next, in \reffis{fig:msmuL-mw} and \ref{fig:msmuR-mw} we show the
dependence of the $\MWMSSM$ prediction on $\msmu{L}$ and $\msmu{R}$,
respectively. For the $\Slpm$-coannihilation case-L naturally a light
$\Smu{L}$ is present in the particle spectrum, close in mass to the
LSP, as can be seen in \reffi{fig:msmuL-mw}.
Consequently, see \reffi{fig:mneu1-mw}, the largest contributions
to $\MWMSSM$ are reached for the lowest values of $\msmu{L}$.
Thus, the low values of $\Smu{L}$ that are required in this scenario in
order to bring the $\MW$ prediction close to the experimental central
value offer
interesting prospects for upcoming searches for SUSY particles.
The case is different for wino DM. Here $\MWMSSM$ values within the
$1\,\sig$ interval of $\MWexp$
are reached for $200 \gev \lsim \msmu{L} \lsim 1200 \gev$, which will
make it difficult to conclusively probe this scenario at the HL-LHC or
the ILC, see also the discussion in \citeres{CHS1,CHS2,CHS3}. Clearly
visible is also a ``hole'' in the $\msmu{L}$ parameter space at
$250 \gev \lsim \msmu{L} \lsim 400 \gev$,
where no scan points are allowed by the applied constraints.
We found that this is due to the experimental constraints from the LHC,
specifically
from slepton pair production searches in the 2l+\met
channel~\cite{ATLAS:2018ojr}.

As shown in \reffi{fig:msmuR-mw}, for the wino DM scenario
an $\MW$ prediction close to the experimental central value is
correlated with a
similar range of $\Smu{R}$ values as it was the case for $\Smu{L}$ in
\reffi{fig:msmuL-mw}. On the other hand,
for the $\Slpm$-coannihilation case-L scenario the $\MW$ prediction is only
weakly sensitive on $\Smu{R}$
in contrast to the case of $\Smu{L}$.
Values of $\MWMSSM$ within the $1\,\sig$ experimental limit are
reached for the whole range of $200 \gev \lsim \msmu{R} \lsim 1300 \gev$.

\begin{figure}[htb!]
\begin{center}
\includegraphics[width=0.9\textwidth]{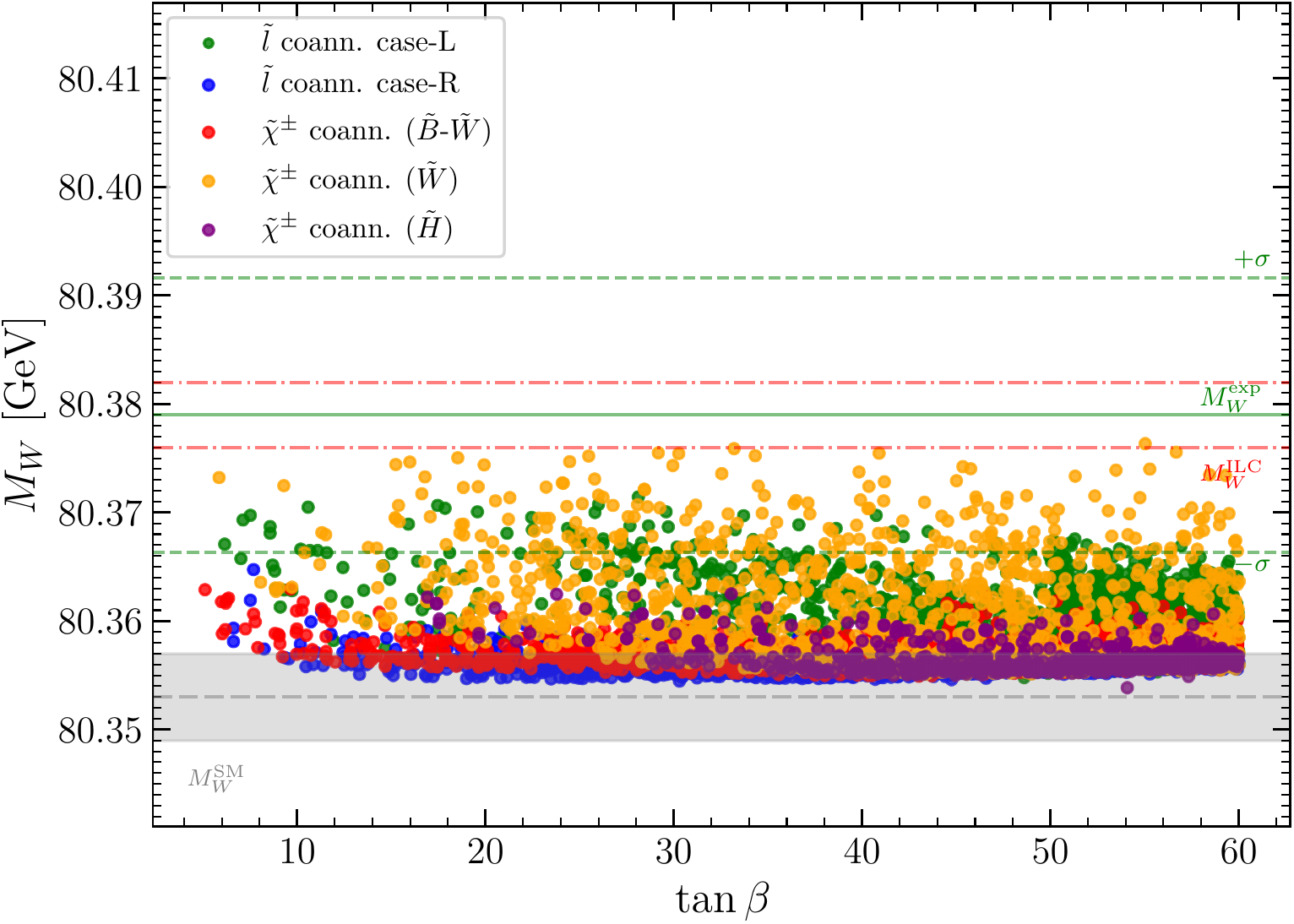}
\caption{\label{fig:tb-mw}
Results for the five scenarios in the $\tb$--$\MW$ plane.
The horizontal lines and the color coding are
as in \protect\reffi{fig:amu-mw}.
}
\end{center}
\end{figure}

As final step of our analysis we show in \reffi{fig:tb-mw} the
dependence of $\MWMSSM$ on $\tb$.
In \citeres{CHS1,CHS2,CHS3} it was
shown that low values of $\mneu1$ can yield a prediction for $\amumssm$
in the preferred region for effectively the whole allowed $\tb$
range. Consequently, no pronounced dependence of $\MWMSSM$ on $\tb$ is
expected, see also \citeres{Heinemeyer:2006px,Heinemeyer:2013dia}. This
is confirmed
in \reffi{fig:tb-mw}. Values of $\MWMSSM$ in the experimental
$1\,\sig$ range are found for $5 \lsim \tb \lsim 60 (50)$ for the wino DM
($\Slpm$-coannihilation case-L) scenario.

%%%%%%%%%%%%%%%%%%%%%%%%%% F I G U R E %%%%%%%%%%%%%%%%%%%%%%%%%%%%%%%%%%%%%%%
%%%%%%%%%%%%%%%%%%%%%%%%%% F I G U R E %%%%%%%%%%%%%%%%%%%%%%%%%%%%%%%%%%%%%%%

%%%%%%%%%%%%%%%%%%%%%%%%%% F I G U R E %%%%%%%%%%%%%%%%%%%%%%%%%%%%%%%%%%%%%%%
%%%%%%%%%%%%%%%%%%%%%%%%%% F I G U R E %%%%%%%%%%%%%%%%%%%%%%%%%%%%%%%%%%%%%%%

%%%%%%%%%%%%%%%%%%%%%%%%%% F I G U R E %%%%%%%%%%%%%%%%%%%%%%%%%%%%%%%%%%%%%%%
%%%%%%%%%%%%%%%%%%%%%%%%%% F I G U R E %%%%%%%%%%%%%%%%%%%%%%%%%%%%%%%%%%%%%%%

%%%%%%%%%%%%%%%%%%%%%%%%%%%%%%%%%%%%%%%%%%%%%%%%%%%%%%%%%%%%%%%%%%%%%%%%%%
%%%%%%%%%%%%%%%%%%%%%%%%%%%%%%%%%%%%%%%%%%%%%%%%%%%%%%%%%%%%%%%%%%%%%%%%%%

\section {Conclusions and outlook}
\label{sec:conclusion}

The new result for the Run~1 data of the ``MUON G-2'' experiment
confirmed the deviation from the SM prediction found previously.
The combination of the experimental results yields a discrepancy with the
theory world average for the SM prediction of
$\Delta\amu = (\newdiff \pm \newdiffunc) \times 10^{-10}$, corresponding
to a $\newdiffsig\,\sig$
effect. Contributions from the EW sector of the MSSM, consisting of charginos,
neutralinos and scalar leptons, can bring the theoretical prediction
into very good agreement with
the new combined average of $\amu^{\rm exp}$, while at the same time
complying with all other experimental and theoretical constraints on this
sector. In particular, identifying
the lightest neutralino, $\neu1$, with the LSP,
a prediction for the CDM relic abundance can be obtained that is in accordance
with the experimental observation, while respecting the bounds from DM
direct detection experiments as well as from direct searches for new particles
at the LHC.

Using the Planck measurement of the DM relic density as an upper limit
on the DM content that is associated with $\neu1$, we analyzed the
prediction for the
$W$-boson mass in the MSSM, $\MWMSSM$. We assumed that the colored sector
of the MSSM is heavy such that it is in agreement with the LHC limits
from direct searches for SUSY particles and at the same time yields a negligible contribution to $\MWMSSM$.
In general, the direct search limits in combination with the requirement
to obtain a prediction for $\Mh$ that within the theoretical uncertainties
is in agreement with the measured mass value
of the SM-like Higgs boson of about $125\gev$ yields a lower bound on
the scale of the stop masses of  $\gsim 1.5 \tev$.
While the squark contributions to $\MWMSSM$ tend to be
relatively small in this mass
region, they are not necessarily negligible if the third generation
squark masses
happen to be close to this lower bound and/or if large mixing is
present. We have checked the numerical impact of
those contributions and found that an upward shift of $\MWMSSM$ of about
$20\mev$ or more is possible for the case of the parameter sets in this study
where the stop and sbottom masses are near
$1.5\tev$ and the constraint on $\Mh$ is satisfied,
see also the discussion in  \citere{Heinemeyer:2013dia}.
Accordingly, an analysis targeting the lowest possible mass values of the third
generation squarks in the MSSM should take into account also the contribution of
the colored sector to $\MWMSSM$. We leave such an investigation for future work.

We analyzed five scenarios, depending on the mechanism that brings the
relic density in agreement with the {observed upper bound}:
bino/wino DM with $\cha1$-coannihilation, bino DM with
$\Slpm$-coannihilation with the mass of the
``left-handed'' (``right-handed'') slepton close to $\mneu1$, case-L
(case-R), wino DM and higgsino DM.
We find that only the scenarios of wino DM and the $\Slpm$-coannihilation
case-L can give rise to sizable contributions to $\MWMSSM$, up to
$\sim 25 \mev$ and $\sim 20 \mev$, respectively.
Accordingly, in those scenarios the predicted values for $\MW$ and $\amu$
can simultaneously be very close to the present experimental central values.
As a consequence, the anticipated future accuracy on $\MW$ offers interesting
prospects for discriminating between different DM scenarios, depending on the
future experimental central value for $\MW$.

For the $\Slpm$-coannihilation case-L the numerically important
corrections stem mostly from the self-energy contributions, whereas for
wino DM the vertex and box contributions are most relevant.
In these scenarios the largest $\MWMSSM$ values are reached for the
smallest experimentally allowed $\neu1$ masses, $\mneu1 \lsim
150 \gev$. Concerning the other EW SUSY masses, sizable contributions
to $\MWMSSM$ require a light chargino, $\mcha1 \lsim 200 \gev$ for
wino DM, or a light ``left-handed'' smuon, $\msmu{L} \lsim 250 \gev$
for the $\Slpm$-coannihilation case-L. The combined analysis of
$\amu$, $\MW$ and the DM relic density therefore leads to preferred mass
regions of the specified SUSY states that can serve as targets
for searches at the (HL-)LHC and future $e^+e^-$~colliders, such as the
``second stage'' ILC with $\sqrt{s} = 500 \gev$.

The complementary information from future searches for new particles and
from an increased sensitivity of the precision observables
$\amu$ and $\MW$ will give rise to stringent tests of the theory of the
electroweak interactions.
We have indicated in our plots the prospective improvement on
the precision for $\MWexp$, displayed for the example of the
expected accuracy at the ILC, $\de\MW^{\rm ILC} \sim 3 \mev$.
If the future central value of $\MWexp$ stays close to the present value,
the SM would be strongly disfavored, and also various MSSM scenarios
would yield a large discrepancy between the theory prediction and the
experimental value of $\MW$. On the other hand, the scenarios with light
EW SUSY particles yielding predictions of $\amu$
and $\MW$ close to the current experimental central values would clearly offer
very good prospects for future $e^+e^-$ colliders.

%%%%%%%%%%%%%%%%%%%%%%%%%%%%%%%%%%%%%%%%%%%%%%%%%%%%%%%%%%%%%%%%%%%%%%%%%%

\subsection*{Note added}
Shortly after submission of this paper a new measurement of $\MW$ was
reported by the CDF collaboration~\cite{CDF:2022hxs}, corresponding to
$\MW^{\rm CDF-new} = 80.4335 \pm 0.0094 \gev$, which lies substantially
above the experimental PDG average, \refeq{mwexp}. The main emphasis of
this article is the correlation between the EW SUSY sector giving a good
description for \gmin2\ and the corresponding effects on the $\MW$
prediction from these EW particles, focusing on the current PDG value of
$\MWexp$ (no new official average including the recent CDF measurement
is available so far). In the future it will be mandatory to assess the
compatibility of the different measurements of $\MW$ and to carefully analyze
possible sources of systematic effects. However, if an
experimental value substantially above the current PDG value (as
indicated by the new CDF measurement) is confirmed, an analysis including the effects of SU(2)~breaking in the
stop/sbottom sector will be necessary, as discussed
in \refse{sec:conclusion}.

%%%%%%%%%%%%%%%%%%%%%%%%%%%%%%%%%%%%%%%%%%%%%%%%%%%%%%%%%%%%%%%%%%%%%%%%%%
%\clearpage

\subsection*{Acknowledgements}

We thank M.~Berger, M.~Falck and G.~Moortgat-Pick for useful discussions.
The work of I.S.\ is supported by World Premier
International Research Center Initiative (WPI), MEXT, Japan.
The work of S.~H.\ is supported in part by the grant
PID2019-110058GB-C21 funded by MCIN/AEI/10.13039/501100011033 and by
"ERDF A way of making Europe", and in part by the grant CEX2020-001007-S
funded by MCIN/AEI/10.13039/501100011033.
The work of M.C.\ is supported by the project AstroCeNT:
Particle Astrophysics Science and Technology Centre,  carried out within
the International Research Agendas programme of
the Foundation for Polish Science financed by the
European Union under the European Regional Development Fund.
G.W.\ acknowledges support by the Deutsche Forschungsgemeinschaft
(DFG, German Research Foundation) under Germany‘s Excellence
Strategy -- EXC 2121 ``Quantum Universe'' – 390833306.

%%%%%%%%%%%%%%%%%%%%%%%%%%%%%%%%%%%%%%%%%%%%%%%%%%%%%%%%%%%%%%%%%%%%%%%%%%
%%%%%%%%%%%%%%%%%%%%%%%%%%%%%%%%%%%%%%%%%%%%%%%%%%%%%%%%%%%%%%%%%%%%%%%%%%

%\clearpage
%\newpage

\newcommand\jnl[1]{\textit{\frenchspacing #1}}
\newcommand\vol[1]{\textbf{#1}}

%%%%% CLEAR DOUBLE PAGE!
\newpage{\pagestyle{empty}\cleardoublepage}

%%%%%%%%%%%%%%%%%%%%%%%%%%%%%%%%%%%%%%%%%%%%%%%%%%%%%%%%%%%%%%%%%%%%%%%%%%%

\end{document}

%% file: paperdef.tex
\newcommand{\lsim}
{\;\raisebox{-.3em}{$\stackrel{\displaystyle <}{\sim}$}\;}
\newcommand{\gsim}
{\;\raisebox{-.3em}{$\stackrel{\displaystyle >}{\sim}$}\;}

\newcommand\al{\alpha}

\newcommand\tb{\tan\beta}

\newcommand\LB{\left[}
\newcommand\RB{\right]}
\newcommand\LV{\left\{}
\newcommand\RV{\right\}}

\newcommand\ReDiag{\mathop{%
  \raise .5pt\hbox{[}%
  \widetilde{\mathrm{Re}}%
  \raise .5pt\hbox{]}}}
\newcommand\ReOffDiag{\mathop{%
  \raise .5pt\hbox{$\llbracket$}%
  \widetilde{\mathrm{Re}}%
  \raise .5pt\hbox{$\rrbracket$}}}

\newcommand\MSbar{\ensuremath{\overline{\mathrm{MS}}}}

\newcommand\MW{M_W}
\newcommand{\MWSM}{\MW^{\SM}}
\newcommand{\MWMSSM}{\MW^{\MSSM}}
\newcommand{\MWexp}{\MW^{\mathrm{exp}}}
\newcommand\MZ{M_Z}
\newcommand\Mh{M_h}
\newcommand\MH{M_H}
\newcommand\MA{M_A}

\newcommand\mt{m_t}

\newcommand\Sl{\tilde l}

\newcommand\Slpm{\tilde l^\pm}

\newcommand\Sel[1]{\tilde e_{#1}}
\newcommand\Smu[1]{\tilde \mu_{#1}}

\newcommand\mse[1]{m_{\Sel{#1}}}
\newcommand\msl[1]{m_{\Sl_{#1}}}

\newcommand\mL{m_{\tilde l_L}}
\newcommand\mR{m_{\tilde l_R}}

\newcommand\ino[1]{\tilde\chi_{#1}}

\newcommand\chapm[1]{\ino{#1}^\pm}

\newcommand\cha{\chapm}
\newcommand\mcha[1]{m_{\chapm{#1}}}

\newcommand\neu[1]{\ino{#1}^0}
\newcommand\mneu[1]{m_{\neu{#1}}}

\newcommand\refeq[1]{Eq.~(\ref{#1})}

\newcommand\refse[1]{Sect.~\ref{#1}}

\newcommand\citere[1]{Ref.~\cite{#1}}
\newcommand\citeres[1]{Refs.~\cite{#1}}

\newcommand{\SM}{\mathrm{SM}}
\newcommand{\MSSM}{\mathrm{MSSM}}

\newcommand{\CP}{{\cal CP}}
\newcommand{\cp}{{\CP}}

\newcommand{\tev}{\,\, \mathrm{TeV}}
\newcommand{\gev}{\,\, \mathrm{GeV}}
\newcommand{\mev}{\,\, \mathrm{MeV}}

\newcommand\edz{\tfrac{1}{2}}

\newcommand\MO{\texttt{MicrOMEGAs}}
\newcommand\CM{\texttt{CheckMATE}}

\newcommand\msmu[1]{m_{\tilde{\mu}_{#1}}}
\newcommand\mstau[1]{m_{\tilde{\tau}_{#1}}}

\newcommand{\sig}{\sigma}

\def\order#1{\ensuremath{{\cal O}(#1)}}
\def\reffi#1{\mbox{Fig.~\ref{#1}}}
\def\reffis#1{\mbox{Figs.~\ref{#1}}}

\def\De{\Delta}
\def\de{\delta}

\def\amumssm{\ensuremath{a_\mu^{\MSSM}}}
\def\gmin2{\ensuremath{(g-2)_\mu}}
\def\amu{\ensuremath{a_\mu}}
\def \met  {\mbox{${E\!\!\!\!/_T}$}}
\newcommand{\ssi}{\ensuremath{\sig_p^{\rm SI}}}

\definecolor{Orange}{named}{orange}
\definecolor{Purple}{named}{purple}
\definecolor{Lightblue}{cmyk}{0.9,0.1,0.1,0.3}
\definecolor{dgelborange}{cmyk}{0.,0.3,0.5, 0.}
\definecolor{Lila}{rgb}{0.5,0.,1}
\definecolor{Darkgreen}{rgb}{0.,.7,0.2}